\documentclass[pre,twocolumn,a4paper,superscriptaddress]{revtex4}
\usepackage[english]{babel}
\usepackage{amsmath}
\usepackage{amsfonts}
\usepackage{amssymb}
\usepackage{array} 
\usepackage{dcolumn}
\usepackage{longtable}
\usepackage{hyperref}
\usepackage{xcolor}
\hypersetup{
    colorlinks,
    linkcolor={red!50!black},
    citecolor={blue!50!black},
    urlcolor={blue!80!black}
}

\usepackage{float}
\usepackage{bbm}
\usepackage{bm}
\usepackage{graphicx}
\usepackage{natbib}
\usepackage{color}
\usepackage{comment}

\definecolor{RevisionColor}{rgb}{0.325, 0.58, 0.722}

\begin{document}
\title{Stress-dependent amplification of active forces in nonlinear elastic media}
\date{\today}
\author{Pierre Ronceray}
\affiliation{Princeton Center for Theoretical Science, Princeton University, Princeton, NJ 08544, USA}
\author{Chase Broedersz}
\affiliation{Arnold-Sommerfeld-Center for Theoretical Physics and Center for NanoScience, Ludwig-Maximilians-Universit\"at M\"unchen, D-80333 M\"unchen, Germany}
\author{Martin Lenz}
\email{martin.lenz@u-psud.fr}
\affiliation{LPTMS, CNRS, Univ. Paris-Sud, Universit\'e Paris-Saclay, 91405 Orsay, France}
\affiliation{MultiScale  Material  Science  for  Energy  and  Environment,
UMI  3466,  CNRS-MIT,
77  Massachusetts  Avenue,
Cambridge,  Massachusetts  02139,
USA}

\begin{abstract}
The production of mechanical stresses in living organisms largely relies on localized, force-generating active units embedded in filamentous matrices. 
Numerical simulations of discrete fiber networks with fixed boundaries have shown that buckling in the matrix dramatically amplifies the resulting active stresses.
Here we extend this result to a bucklable continuum elastic medium subjected to an arbitrary external stress, and derive analytical expressions for the active, nonlinear constitutive relations characterizing the full active medium.
Inserting these relations into popular ``active gel'' descriptions of living tissues and the cytoskeleton will enable investigations into nonlinear regimes previously inaccessible due to the phenomenological nature of these theories.
\end{abstract}

\maketitle

\section{Introduction}

Cells move and deform in response to stresses. These stresses
originate both from the deformation of their environment, and from the
active forces they generate internally.  Within the cell, these forces
are largely generated by molecular motors acting at the nanometer
scale that are embedded in a matrix of semiflexible filaments known as
the actin cytoskeleton. The cytoskeleton then transmits these forces
to larger length scales, allowing them to control shape and generate
stresses over the whole cell. At even larger length scales, the
resulting cell-wide forces can be further transmitted by another type
of fibrous network, the extracellular matrix, and this transmission
results in stress production over several millimeters in connective
tissues~\cite{Jen:1982}. Much progress has been made recently in
understanding how these active forces are transmitted by fiber
networks from the microscopic to macroscopic scales, thus enabling cell motion and division, wound healing or embryonic development~\cite{Blanchoin:2014,Heisenberg:2013}. Furthermore, it
is now well understood how passive biopolymer networks, both inside
and outside cells, respond to external strain~\cite{Broedersz:2014}. However, little
is known about the interplay between internal stress generation and
external stresses due to environment strain.

\begin{figure}
\centering
\includegraphics[width=\columnwidth]{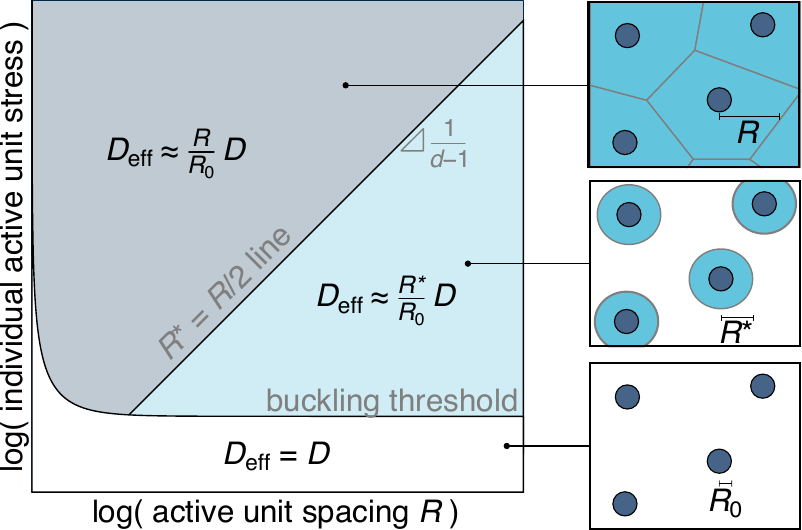}
\caption{\label{fig:summary}A medium populated by strong enough contractile active units (dark blue) buckles and amplifies their active stresses. The phase diagram on the left delimits three buckling regimes, whose physical structures are illustrated by panels on the right. White regime:
Weak active units do not induce buckling, yielding a far-field stress given by Eq.~\eqref{eq:introlinearactivestress}. Blue regime: stronger active units locally buckle the network by exerting compressive stresses in excess of the buckling threshold $\sigma_b$. This endows each active unit with a larger effective dipole $D_\text{eff}\approx (R^*/R_0)D$, where $R^*$ is the radius of the light blue buckled region and $R_0$ that of an active unit. Grey regime: For a medium with fixed boundary, the buckling radius asymptotically goes to the distance $R$ between active units as the strength of the active units becomes very large, implying that $D_\text{eff}\approx (R/R_0)D$. Here we show that imposing an external stress $\sigma$ at the boundary of the medium modifies the values of $D_\text{eff}$ and $R^*$ in the same way as a shift of the buckling threshold from $\sigma_b$ to $\sigma_b+\sigma$.
}
\end{figure}

The key to a theory of stress generation in fiber networks is
understanding how they transmit forces from small to large scales.  While the quantitative
relationship between microscopic forces and the resulting macroscopic
stresses is remarkably simple in linear elastic
media~\cite{Gurtin:1972,Ronceray:2015}, this force transmission is
drastically modified by the nonlinear response conferred to fibrous
media by the buckling of their
filaments~\cite{Shokef:2012,Wang:2014,Notbohm:2014,Rosakis:2014,Xu:2015,Ronceray:2016}. Quantitatively,
there the tensile stress $\sigma_\text{active}$ actively generated by
a density $\rho$ of active units each exerting a force dipole $D$ can
exceed the linear prediction
\begin{equation}\label{eq:introlinearactivestress}
\sigma_\text{active}^\text{(lin)}=-\rho D
\end{equation}
by several orders of
magnitude. Qualitatively, strong active units locally deform the
networks and thus surround themselves with a potentially large buckled region,
where the network is mechanically equivalent to a collection of tense
radial ropes. 
Such stress propagation patterns are described by the general mathematical formalism of tension field theory~\cite{Wagner:1929}, and are also encountered in thin, easily buckled elastic sheets~\cite{Davidovitch:2011}.
As the ropes transmit the forces produced by the active
unit to the boundary of that buckled region, the system comprised of
the active unit plus the ropes acts like an enlarged, effective force
dipole. This effective dipole has an enhanced span compared to the
original one, and thus a larger magnitude $|D^\text{eff}|>|D|$ (Fig.~\ref{fig:summary}).
However, how external strain affects stress generation and modifies
these scaling laws is not known.  Moreover, a detailed analytical
understanding of buckling-induced stress amplification is missing,
although other types of nonlinearities have been investigated in two
dimensions~\cite{Shokef:2012,Xu:2015}.

In this paper, we demonstrate that the effect of external stress on
active stress generation can be simply understood as an enhancement of
the buckling threshold.  To this aim, we derive a full analytical
description of active stress amplification in a simple model of
bucklable medium subjected to an arbitrary external isotropic stress
in any dimension. We restrict our study to isotropic, contractile
active units, motivated by the observation that they represent the
generic far-field response of a fiber network to any large local force
dipole, be it locally contractile or extensile, isotropic or
anisotropic~\cite{Ronceray:2016}. We present the ingredients of our
model in Sec.~\ref{sec:model}, and compute the characteristics of the
forces transmitted by our active medium in
Sec.~\ref{sec:transitions}. We then deduce the resulting macroscopic
stresses in Sec.~\ref{sec:amplification}, and use these expressions to
derive constitutive stress-strain relations for the active medium in
Sec.~\ref{sec:constitutive}. Finally, we discuss our results in
Sec.~\ref{sec:discussion}.

The analytical expressions derived here are key to incorporating the
wealth of available biological and mechanical information about
individual active units in so-called active gel theories, which are
widely used theoretical descriptions of living tissues and the
cytoskeleton~\cite{Kruse:2005aa,Julicher:2007,Joanny:2009}. Indeed,
such theories typically adopt a purely macroscopic point of view, and
while active stresses are the fundamental drivers of the new physics
they explore, active gel descriptions typically assume them to be
constant for lack of a better description~\cite{Prost:2015}.

\section{\label{sec:model}Model}
Our aim is to model a fiber network subject to stresses that are both and externally applied and induced internally by active units. To this end, we consider a homogeneous nonlinear medium in spatial dimension $d$ (with $d=2$ or $3$ in practice) subjected to an isotropic external stress $\sigma$, and within which a density $\rho$ of active units are embedded. Assuming for simplicity that the active units are positioned on a regular lattice (\emph{e.g.}, a triangular lattice in 2D), we focus on the Voronoi cell surrounding one of the active units (\emph{e.g.}, a hexagon in a triangular lattice). We further approximate this cell by a spherical domain with the same volume of as the Voronoi cell, allowing us to consider only spherically symmetric configurations in the following. The radius $R$ of this sphere as a function of the motor density is set by $\rho S_{d-1}R^d=1$, with $S_n$ the volume of the unit $n$-sphere ($S_1=\pi$, $S_2=4\pi/3$).

To account for fiber buckling, the  continuum elastic medium can locally buckle when compressed beyond a critical stress $\sigma_b$. To implement this feature in the simplest fashion, we assume that the medium responds linearly with Lam\'e coefficients $\lambda$ and $\mu$, but that compressive stresses saturate beyond the threshold value $-\sigma_b$. To express this relation formally, we  denote the strain and stress tensors by $\boldsymbol{u}$ and $\boldsymbol{\sigma}$ respectively, and note that the spherical symmetry of the system imposes that both tensors take a diagonal form in spherical coordinates, resulting in the following block structure:
\begin{equation}
\boldsymbol{u}=\left(
\begin{array}{cc}
u_{rr} & 0\\
0 & u_{\theta\theta}I
\end{array}
\right)
\quad \text{and} \quad
\boldsymbol{\sigma}=\left(
\begin{array}{cc}
\sigma_{rr} & 0\\
0 & \sigma_{\theta\theta}I
\end{array}
\right),
\end{equation}
where $I$ is the $(d-1)$-dimensional unit matrix.
In this simple geometry, the radial and orthoradial stresses in the linear regime read
\begin{subequations}\label{eq:linearconstitutive}
\begin{align}
\sigma_{rr}^\text{lin} & = (\lambda+2\mu)u_{rr}+(d-1)\lambda u_{\theta\theta}\\
\sigma_{\theta\theta}^\text{lin} & =
\lambda u_{rr}+[(d-1)\lambda+2\mu] u_{\theta\theta},
\end{align}
\end{subequations}
and our buckling condition can be formulated as
\begin{align}
&\text{if $\sigma_{rr}^\text{lin} > -\sigma_b$ and $\sigma_{\theta\theta}^\text{lin} > -\sigma_b$, } \text{then }\left\lbrace
\begin{array}{l}
\sigma_{rr}=\sigma_{rr}^\text{lin}\nonumber\\
\sigma_{\theta\theta}=\sigma_{\theta\theta}^\text{lin}
\end{array}\right.\label{eq:constitutive}\\
&\text{if $\sigma_{rr}^\text{lin} < -\sigma_b$, } \text{then }\sigma_{rr}=-\sigma_b\\
&\text{if $\sigma_{\theta\theta}^\text{lin} < -\sigma_b$, } \text{then }\sigma_{\theta\theta}=-\sigma_b.\nonumber
\end{align}
Note that we do not need to make specific assumptions about the strain dependence of $\sigma_{rr}$ when $\sigma_{\theta\theta}^\text{lin} < -\sigma_b$ (or that of $\sigma_{\theta\theta}$ when $\sigma_{rr}^\text{lin} < -\sigma_b$) for the purpose of this study, since these components of the stress are then fully determined by force balance.

\begin{figure}[t]
\centering
\includegraphics[width=\columnwidth]{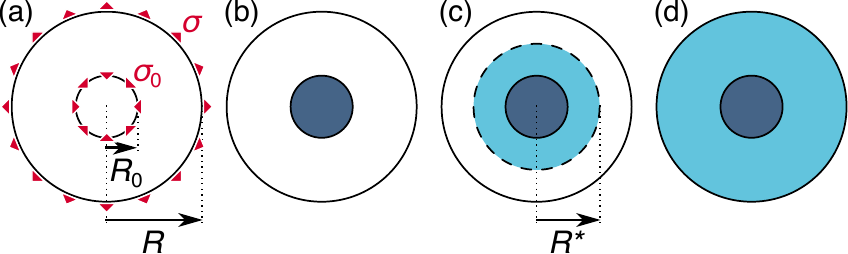}
\caption{\label{fig:bucklingstates}
Successive buckling states of the elastic medium. Red arrowheads on the first panel picture the active internal compression applied across the $r=R_0$ circle, as well as the tensile boundary stress $\sigma$. White regions are in the linear regime in both the radial and orthoradial directions in the sense of Eq.~\eqref{eq:constitutive}. Light blue regions are buckled in the orthoradial directions only, and dark blue regions are buckled in both directions. Qualitatively, as the active unit generates a strong local compressive stress at the center of the medium, buckling is initiated there, then progresses outwards as the active unit stress $\sigma_0$ is increased. A large value of the tensile prestress $\sigma$ antagonizes this compression, delays buckling and hinders the amplification of active stresses.
}
\end{figure}

The elastic medium is centred around an active unit, consisting of a sphere of radius $R_0<R$ imposing a contractile stress $\sigma_0>0$ [Fig.~\ref{fig:bucklingstates}(a)]. This geometry yields a stress discontinuity at the surface of the active unit
\begin{equation}\label{eq:activeunitstress}
\lim_{\epsilon\rightarrow 0}\left[\sigma_{rr}(R_0+\epsilon)-\sigma_{rr}(R_0-\epsilon)\right]=\sigma_0.
\end{equation}
Defining the force dipole exerted by a spatial distributions $f_i(\mathbf{r})$ of body forces as $D_{ij}=\int r_if_j(\mathbf{r})\,\text{d}\mathbf{r}$, our spherically symmetric dipole reads $D_{ij}=D\delta_{ij}$ with $D=-S_{d-1}\sigma_0 R_0^d$. We further assume that the elastic medium is held under constant stress $\sigma_{ij}=\sigma\delta_{ij}$ at its outer boundary, implying the boundary condition
\begin{equation}\label{eq:boundarystress}
\sigma_{rr}(R)=\sigma,
\end{equation}
where this external stress $\sigma$ may be positive or negative.

To compute the stress and displacements associated with our active unit, we must solve the mechanical equilibrium equations $\nabla_j\sigma_{ij}=0$ for our elastic medium, which in our spherical geometry reads
\begin{equation}\label{eq:forcebalance}
\frac{1}{r^{d-1}}\frac{\text{d}(r^{d-1}\sigma_{rr})}{\text{d}r}-\frac{(d-1)\sigma_{\theta\theta}}{r}=0,
\end{equation}
where $\sigma_{rr}$ and $\sigma_{\theta\theta}$ are related to the strain by Eq.~\eqref{eq:constitutive}. We can furthermore express the strain as a function of the radial displacement $u(r)$ of the elastic medium through
\begin{subequations}\label{eq:displacement}
\begin{align}
u_{rr}(r)&=\text{d}u/\text{d}r\\
u_{\theta\theta}(r)&=u/r.
\end{align}
\end{subequations}
Finally, $u(0)=0$ due to spherical symmetry.

\section{\label{sec:transitions}Buckling transitions}
Depending on the values of $\sigma$ and $\sigma_0$, our elastic medium undergoes a sequence of buckling transitions, as illustrated in Fig.~\ref{fig:bucklingstates}. In the following we completely characterize this sequence for non-auxetic materials, \emph{i.e.} materials with a positive Poisson ratio, or equivalently a positive $\lambda$.

For low active unit stresses $\sigma_0$, the material responds linearly [Fig.~\ref{fig:bucklingstates}(a)], and we supplement Eqs.~(\ref{eq:linearconstitutive}-\ref{eq:displacement}) with the requirement that $u(r)$ be continuous in $R_0$. This yields
\begin{subequations}\label{eq:linear}
\begin{align}
u(r<R_0)=&\left\lbrace
\frac{\sigma}{d\lambda+2\mu}-\frac{\sigma_0}{d(\lambda+2\mu)}\right.\nonumber\\
&\times\left.\left[1+\frac{2\mu(d-1)}{d\lambda+2\mu}\left(\frac{R_0}{R}\right)^d\right]
\right\rbrace r\\
u(r>R_0)=&\left[
\frac{\sigma}{d\lambda+2\mu}-\frac{2\mu(d-1)\sigma_0}{d(\lambda+2\mu)(d\lambda+2\mu)}\left(\frac{R_0}{R}\right)^d
\right]r\nonumber\\
&-\frac{\sigma_0}{d(\lambda+2\mu)}\frac{R_0^d}{r^{d-1}}.
\end{align}
\end{subequations}

Increasing the active unit strength $\sigma_0$ from this linear regime puts the $r<R_0$ region under an increasing isotropic compressive stress. As this compressive stress reaches the $-\sigma_b$ threshold for $\sigma_0=\sigma_0^\text{buckling 1}$, with
\begin{equation}
\sigma_0^\text{buckling 1}=\frac{d(\lambda+2\mu)(\sigma+\sigma_b)}{d\lambda+2\mu+2\mu(d-1)(R_0/R)^d},
\end{equation}
the buckling regime of Fig.~\ref{fig:bucklingstates}(b) sets in. In the central buckled region, Eq.~\eqref{eq:constitutive} then implies $\sigma_{rr}(r<r_0)=\sigma_{\theta\theta}(r<r_0)=-\sigma_b$.
Further solving Eqs.~(\ref{eq:linearconstitutive}-\ref{eq:displacement}) in the region $R_0<r<R$ where the medium responds linearly, we find
\begin{align}\label{eq:innerbucklingdisplacement}
u(r>R_0)=&\left[\frac{\sigma}{1-(R_0/R)^d}+\frac{\sigma_b-\sigma_0}{(R/R_0)^d-1}\right]\frac{r}{d\lambda+2\mu}\nonumber\\
&+\frac{\sigma+\sigma_b-\sigma_0}{2\mu(d-1)(R_0^{-d}-R^{-d})}\frac{1}{r^{d-1}}.
\end{align}

\begin{figure}
\centering
\includegraphics[width=\columnwidth]{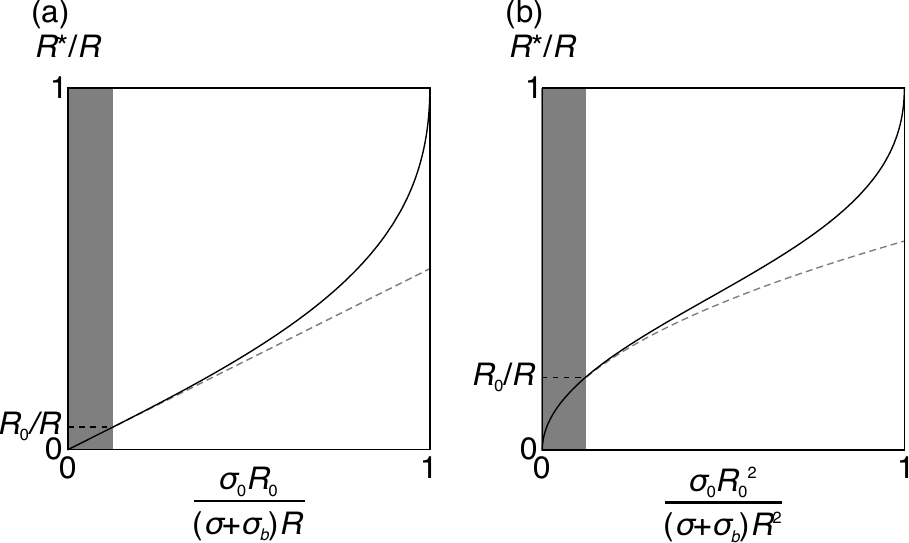}
\caption{\label{fig:bucklingradii}Buckling radii characterizing the buckling regime of Fig.~\ref{fig:bucklingstates}(c) in (a)~$d=2$ and (b)~$d=3$ as given by Eq.~\eqref{eq:bucklingradii} (solid black line) along with the asymptotic expression of Eq.~\eqref{eq:approxbucklingradii} (dashed gray line). The region of the curve corresponding to $R^*<R_0$ (gray box) is not relevant, as it corresponds to an active unit stress $\sigma_0<\sigma_0^\text{buckling 2}$, and thus to another buckling regime. Changes in the value of $R_0/R$ result in a displacement of the boundary of the gray box, while the black curve is unaffected.
}
\end{figure}

Upon a further increase of the active unit strength, the compressive orthoradial stress $\sigma_{\theta\theta}(R_0^+)$ at the outer surface of the active unit reaches the buckling threshold for $\sigma_0=\sigma_0^\text{buckling 2}$, with
\begin{equation}\label{eq:buckling2}
\sigma_0^\text{buckling 2}=\frac{d(\sigma+\sigma_b)}{1+(d-1)(R_0/R)^d}.
\end{equation}
Beyond this threshold, the elastic medium buckles in the orthoradial direction in the region outside the active unit, as pictured in Fig.~\ref{fig:bucklingstates}(c). We denote by $R^*$ the outer limit of this buckling zone. In this regime, $\sigma_{rr}(r<R_0)=\sigma_{\theta\theta}(r<R^*)=-\sigma_b$. Radial force balance additionally imposes that $\sigma_{rr}$ be a continuous function in $R^*$, and the value of the buckling radius $R^*$ is set by the buckling condition $\sigma_{\theta\theta}(R^*)=-\sigma_b$. Solving Eqs.~(\ref{eq:linearconstitutive}-\ref{eq:displacement}) while taking into account these new boundary conditions yields
\begin{subequations}\label{eq:bucklingregion}
\begin{align}
\sigma_{rr}(R_0<r<R^*)= & \sigma_0\left(\frac{R_0}{r}\right)^{d-1}-\sigma_b\label{eq:slowdecay}\\
u(r>R^*)=&\frac{\sigma(R/R^*)^d+\sigma_b-\sigma_0(R_0/R^*)^{d-1}}{(d\lambda+2\mu)[(R/R^*)^d-1]}r\nonumber\\
&+\frac{\sigma+\sigma_b-\sigma_0(R_0/R^*)^{d-1}}{2\mu(d-1)[(R^*)^{-d}-R^{-d}]}\frac{1}{r^{d-1}},\label{eq:bucklingregiondisplacement}
\end{align}
\end{subequations}
where $R^*$ is the solution of the following equation:
\begin{equation}\label{eq:bucklingradii}
\left(\frac{R^*}{R}\right)^d-\frac{d}{d-1}\frac{\sigma+\sigma_b}{\sigma_0}\left(\frac{R^*}{R_0}\right)^{d-1}+\frac{1}{d-1}=0.
\end{equation}
Equation~\eqref{eq:bucklingradii} implies that as long as the buckling zone is much smaller than the size of the entire system ($R^*\ll R$) its radius is given by
\begin{equation}\label{eq:approxbucklingradii}
R^*=R_0\left[\frac{\sigma_0}{d(\sigma+\sigma_b)}\right]^{1/(d-1)},
\end{equation}
which confirms the scaling postulated in Ref.~\cite{Ronceray:2016}. More broadly, in $d=2$ the buckling radius is given by
\begin{equation}
R^*/R=\frac{(\sigma+\sigma_b)R}{\sigma_0R_0}-\sqrt{\left[\frac{(\sigma+\sigma_b)R}{\sigma_0R_0}\right]^2-1},
\end{equation}
while the $d=3$ solution can also be expressed in a closed analytical form, albeit a cumbersome one. We plot both of these solutions in Fig.~\ref{fig:bucklingradii}.
Equation~(\ref{eq:slowdecay}) confirms the observation made in Ref.~\cite{Ronceray:2016} that radial stresses decay slowly with a $r^{1-d}$ power law within the $R_0<r<R^*$ buckling region, thus accounting for long-range stress transmission in buckled systems, in contrast with the $r^{-d}$ decay characteristic of linear materials.
Throughout the regime described here, the buckling zone is under strong radial tensile stress, while it is essentially crumpled in the orthoradial direction, implying that $\sigma_{\theta\theta}$ provides little help in stabilizing the system against the radial tension. As a result, the buckling zone is prevented from collapsing primarily by the unbuckled shell surrounding it, which we picture in white in Fig.~\ref{fig:bucklingstates}(c). 

As the active unit stress $\sigma_0$ is increased yet again, the buckling radius $R^*$ reaches the boundary of the system for $\sigma_0=\sigma_0^\text{collapse}$, with
\begin{equation}
\sigma_0^\text{collapse}=(\sigma+\sigma_b)\left(\frac{R}{R_0}\right)^{d-1}.
\end{equation}
Beyond this value, the stabilizing unbuckled outer shell vanishes and the system collapses. Formally, this collapse is manifested by  the mechanical equilibrium equation Eq.~\eqref{eq:forcebalance} having no solution that satisfies both boundary conditions Eqs.~\eqref{eq:activeunitstress} and \eqref{eq:boundarystress}.

\begin{figure}
\centering
\includegraphics[width=\columnwidth]{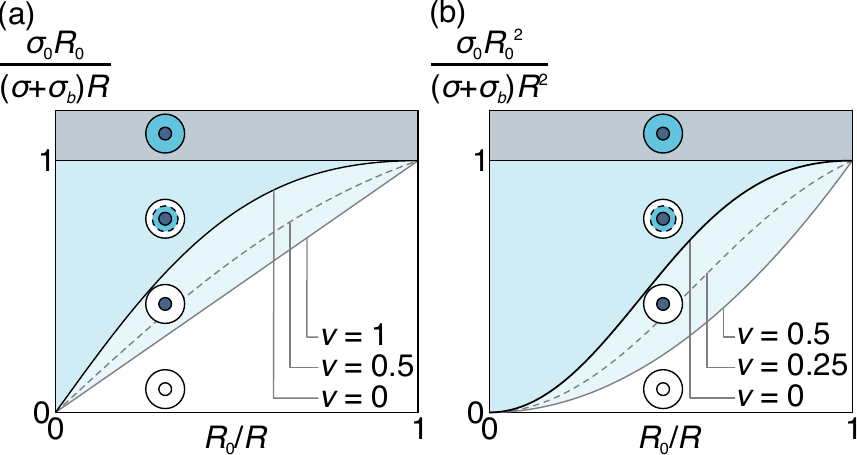}
\caption{\label{fig:phasediag}Parameter values characterizing the four possible buckling regimes in dimensions (a)~$d=2$ and (b)~$d=3$. The lines on the diagrams are the boundaries of the different regimes. The position of the lower line (in grey) depends on the Poisson modulus $\nu=(d-1+2\mu/\lambda)^{-1}$ of the material, and its position is indicated for three values of $\nu$ on each panel. In particular, for $\nu=0$ this line is identical to the central black line and the second buckling regime of Fig.~\ref{fig:bucklingstates} is thus nonexistent. Note that thermodynamic stability requires $\nu\leqslant 1$ in $d=2$ and $\nu\leqslant 0.5$ in $d=3$.
}
\end{figure}

We illustrate the parameter ranges associated with the four buckling regimes discussed in this section in the phase diagram of Fig.~\ref{fig:phasediag}.

\section{\label{sec:amplification}Stress amplification}
The external stress $\sigma$ applied at the boundary of the elastic medium is balanced by two contributions: a passive elastic response of the network, and an active stress specifically due to the presence of these active units
\begin{equation}\label{eq:stressdecomposition}
\sigma=\sigma_\text{elastic}+\sigma_\text{active},
\end{equation}
This decomposition of total stress into a passive and an active contribution is a central ingredient of active gel theories~\cite{Kruse:2005aa,Ronceray:2015}, where the contribution of $\sigma_\text{active}$ drives nonequilibrium flows and pattern formation \cite{Voituriez:2005,Julicher:2007,Joanny:2009,Bois:2011,Kumar:2014}. To determine $\sigma_\text{active}$, we determine $\sigma_\text{elastic}$ as the stress that would be required to impose the same boundary displacement observed in our system onto a purely passive, $\sigma_0=0$ medium. Thus,
\begin{equation}\label{eq:elasticstress}
\sigma_\text{elastic}=(d\lambda+2\mu)\frac{u(R)}{R}
\end{equation}
Note that the previously studied special case of a fixed boundary corresponds to $\sigma_\text{elastic}=0$~\cite{Ronceray:2016}. Here we combine the displacements computed in Sec.~\ref{sec:transitions} with Eqs.~(\ref{eq:stressdecomposition}-\ref{eq:elasticstress}) to compute the dependence of the two stress contributions on the parameters of our model.

For a completely homogeneous linear (but possibly anisotropic) network, general linear elastic considerations~\cite{Gurtin:1972,Ronceray:2015} impose that the active stress is proportional to the force dipole and density of active units through $\sigma_\text{active}^\text{(lin)}=-\rho D$. Indeed, combining Eq.~\eqref{eq:linear} with Eqs.~(\ref{eq:stressdecomposition}-\ref{eq:elasticstress}) yields
\begin{equation}\label{eq:linearactivestress}
\sigma_\text{active}^\text{(lin)}=\sigma_0\left(\frac{R_0}{R}\right)^d=-\rho D.
\end{equation}

When active units are strong enough to buckle the network, the reference active stress of Eq.~\eqref{eq:linearactivestress} is amplified, which we quantify through the amplification factor
\begin{equation}
{\cal A}=\frac{\sigma^\text{active}}{\sigma_\text{active}^\text{(lin)}}.
\end{equation}
Combining the displacements of Eqs.~\eqref{eq:innerbucklingdisplacement} and \eqref{eq:bucklingregiondisplacement} with Eqs.~(\ref{eq:stressdecomposition}-\ref{eq:elasticstress}) thus yields
\begin{equation}
{\cal A}=\frac{(\lambda+2\mu)d}{2\mu(d-1)}\frac{1}{1-(R_0/R)^d}\left(1-\frac{\sigma+\sigma_b}{\sigma_0}\right)
\end{equation}
for the inner buckling regime illustrated in Fig.~\ref{fig:bucklingstates}(b) and
\begin{equation}\label{eq:bucklingregionactivestress}
{\cal A}=\left(1+\frac{\lambda}{2\mu}\right)\frac{R^*}{R_0}
\end{equation}
for the regime of Fig.~\ref{fig:bucklingstates}(c). This last relationship validates the ${\cal A}\propto R^*/R_0$ scaling postulated for the ``force-controlled'' regime of Ref.~\cite{Ronceray:2016} on the basis of the amplified force dipole picture described in the introduction. Finally, no amplification factor can be computed for the collapsing regime of Fig.~\ref{fig:bucklingstates}(d) as it does not give rise to a well-defined boundary displacement $u(R)$. Fully buckled systems can, however, be realized in systems with fixed boundaries preventing this collapse; in such systems, the fixed boundary imposes a stress $\sigma$ satisfying Eq.~\eqref{eq:buckling2} that maintains the system at the threshold between the regimes of Figs.~\ref{fig:bucklingstates}(c) and (d). This sets the amplification at the $R^*\rightarrow R$ limit of Eq.~(\ref{eq:bucklingregionactivestress}):
\begin{equation}\label{eq:saturatedactivestress}
 {\cal A}=\left(1+\frac{\lambda}{2\mu}\right)\frac{R}{R_0}, 
 \end{equation}
and corresponds to the ``density-controlled'' regime of Ref.~\cite{Ronceray:2016}.

Equations~(\ref{eq:linearactivestress}-\ref{eq:saturatedactivestress}) constitute a complete description of the active stress produced by the system as a function of the linear moduli and buckling stresses of the elastic medium, as well as the density and strength of the active units. These active stresses depend on the externally applied stress $\sigma$ in the buckled regimes, as tensing the medium antagonizes buckling and amplification, which in turn decreases the active stress.

\section{\label{sec:constitutive}Constitutive relations}
Constitutive stress-strain relations for the active material can be derived from Eqs.~(\ref{eq:linearactivestress}-\ref{eq:saturatedactivestress}). Denoting the isotropic strain by $\gamma=u(R)/R$, we find that in the linear regime  Eq.~\eqref{eq:stressdecomposition} can be rewritten as
\begin{subequations}\label{eq:LinearStressStrain}
\begin{equation}
\sigma=(d\lambda+2\mu)\gamma+\sigma_0\left(\frac{R_0}{R}\right)^d
\end{equation}
\emph{i.e.}, an affine stress-strain relation involving the same elastic modulus as for the passive system, consistent with the most common formulations of active gels theories. This linear regime is valid for large enough strains, namely
\begin{equation}
\gamma+\gamma_b>\frac{1-(R_0/R)^d}{\lambda+2\mu}\frac{\sigma_0}{d},
\end{equation}
\end{subequations}
where $\gamma_b=\sigma_b/(\lambda d+2\mu)$ is the absolute value of the critical strain at which the elastic medium buckles in the absence of active units.
At lower strains, the buckling regime of Fig.~\ref{fig:bucklingstates}(b) takes over and yields a different stress-strain relation, albeit still an affine one:
\begin{subequations}
\begin{align}
\sigma=&\frac{2\mu(d-1)(d\lambda+2\mu)[1-(R/R_0)^d]}{2\mu(d-1)+(d\lambda+2\mu)(R/R_0)^d}\gamma\nonumber\\
&+\frac{d(\lambda+2\mu)(R/R_0)^d}{2\mu(d-1)+(d\lambda+2\mu)(R/R_0)^d}(\sigma_0-\sigma_b)
\end{align}
valid for strains
\begin{equation}
\left[\frac{1}{\lambda d+2\mu}-\frac{(R_0/R)^d}{2\mu}\right]\frac{\sigma_0}{d}<\gamma+\gamma_b<\frac{1-(R_0/R)^d}{\lambda+2\mu}\frac{\sigma_0}{d}
\end{equation}
\end{subequations}
Finally, for even lower (or more compressive) strains the buckling zone regime of Fig.~\ref{fig:bucklingstates}(c) takes over, and the stress is nonlinearly related to the strain through
\begin{equation}\label{eq:bucklingzoneconstitutive}
\sigma = (d\lambda+2\mu)\gamma+\left(1+\frac{\lambda}{2\mu}\right)\sigma_0\frac{R_0^{d-1}R^*}{R^d}, 
\end{equation}
where $R^*$ itself is a function of $\sigma$ through Eq.~\eqref{eq:bucklingradii}. Inserting Eq.~\eqref{eq:bucklingzoneconstitutive} into Eq.~\eqref{eq:bucklingradii} and defining
\begin{subequations}
\begin{align}
\tilde{\sigma}&=\frac{\sigma+\sigma_b}{\sigma_0}\\
\tilde{\gamma}&=\frac{2\mu(\gamma+\gamma_b)}{\sigma_0},
\end{align}
\end{subequations}
we get the following relation between stress and strain
\begin{equation}
\left[\tilde{\sigma}-\left(1+\frac{d\lambda}{2\mu}\right)\tilde{\gamma}\right]^{d-1}\left[\tilde{\sigma}+(d-1)\tilde{\gamma}\right]=\frac{(1+\lambda/2\mu)^d}{1+d\lambda/2\mu}\left(\frac{R_0}{R}\right)^{d(d-1)}
\end{equation}
which is a polynomial equation of order $d$ in $\sigma$ and can be solved for $\sigma$ in both $d=2$ and $d=3$. Here we present the more compact $d=2$ result:
\begin{subequations}\label{eq:bucklingzonestressstrain}
\begin{equation}
\tilde{\sigma}=
\frac{\lambda\tilde{\gamma}}{\mu}
+\left(1+\frac{\lambda}{2\mu}\right)\sqrt{\frac{(R_0/R)^2}{1+\lambda/\mu}+\tilde{\gamma}^2},
\end{equation}
and we give the bounds of this buckling zone regime in arbitrary dimension:
\begin{equation}\label{eq:bucklingzonestraininterval}
-\frac{\lambda(R_0/R)^{d-1}}{2\mu(d\lambda+2\mu)}\sigma_0<\gamma+\gamma_b<\left[\frac{1}{\lambda d+2\mu}-\frac{(R_0/R)^d}{2\mu}\right]\frac{\sigma_0}{d},
\end{equation}
\end{subequations}
where the lower bound of Eq.~\eqref{eq:bucklingzonestraininterval} represents the critical strain for the transition to the collapsed state of Fig.~\ref{fig:bucklingstates}(d).

\begin{figure}
\centering
\includegraphics[width=\columnwidth]{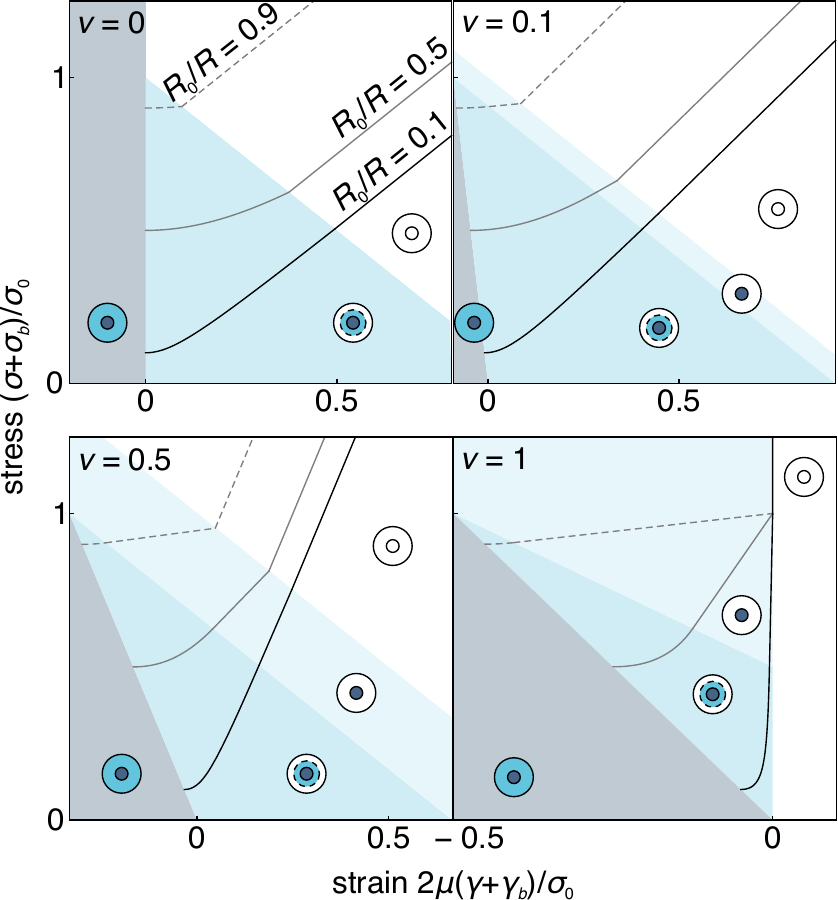}
\caption{\label{fig:StressStrain}Stress-strain relations for a $d=2$ medium with embedded active units as a function of the Poisson ratio of the material (indicated in the top left corner of each panel) and the size of the active unit (see labeling of the lines in the top left panel). The colored regions denote the buckling regimes of Fig.~\ref{fig:bucklingstates}. The slope of the stress-strain curves always vanishes as the system becomes unstable at the boundary of the collapsed (dark grey) regime.}
\end{figure}

Equations~(\ref{eq:LinearStressStrain}-\ref{eq:bucklingzonestressstrain}) form a complete nonlinear constitutive relation relating the stress $\sigma$ to the strain $\gamma=u(R)/R$ in elastic systems with embedded active units. We illustrate this relation in Fig.~\eqref{fig:StressStrain}, which shows that the resulting active material always softens under compression before losing stability as the collapsing threshold is reached at low enough $\gamma$. In addition, the influence of the material's buckling threshold $\sigma_b=(d\lambda+2\mu)\gamma_b$ and active unit stress $\sigma_0$ on this relation is remarkably simple, as they respectively result in a shift and a rescaling in the values of the stress and strain.

\section{\label{sec:discussion}Discussion}

Active units embedded in fibrous media, such as molecular motors or
whole contractile cells, exert strong forces on their
surroundings. These active forces deform and buckle the fibers, thus
affecting the way in which these forces are transmitted.  Here we
present a detailed analysis of this process and provide constitutive
relations describing the material properties emerging from
interactions between active unit and fiber networks. Such relations
can readily be incorporated into macroscopic descriptions of active
systems~\cite{Prost:2015}. At the microscopic length scale, they could
conversely be supplemented with more detailed dynamical descriptions
of the way in which the active unit stress $\sigma_0$ is
produced~\cite{Chan:2008,Lenz:2014} to elucidate the coupled dynamics
of an active unit and its elastic environment.

Our present results show that buckling in fiber networks results in an
amplification of active stress. The buckling transitions underlying
the force transmission described here proceed in several steps. The
first step involves the buckling of the system's core shown in
Fig.~\ref{fig:bucklingstates}(b). This regime is clearly tied to our
specific description of the active unit as a sphere of radius $R_0$, and
may be substantially modified when using active units with different
geometries. In the second buckling regime, a potentially large region
surrounding of the active unit undergoes orthoradial buckling
[Fig.~\ref{fig:bucklingstates}(c)]. Contrary to the previous one, we
expect this regime to be largely insensitive to the details of the
active units, as the nonlinear response of fiber networks gives rise
to an emergent isotropic force dipole away from the active
units~\cite{Ronceray:2016}. In the case of sparse active units
$R_0\ll R$, this regime occupies a much larger fraction of parameter
space than the previous one (see Fig.~\ref{fig:phasediag}), implying
that its universal physics dominates nonlinear force transmission in
systems with a low volume fraction of active units. Finally, the last
transition considered here [Fig.~\ref{fig:bucklingstates}(d)]
corresponds to the limit of stability of the system, which cannot be
described in a fixed stress ensemble. Indeed, the network does not
have an intrinsic shape anymore, and collapses if its boundaries are
released. However, at fixed boundary strain (\emph{e.g.} for fixed or
periodic boundary conditions), the system is characterized by a
well-defined active stress, with an amplification factor proportional to $R/R_0$, as
predicted numerically in our previous work~\cite{Ronceray:2016}.

Our findings confirm and extend several heuristic conclusions
formulated in our previous scaling arguments and numerical simulations
of explicit filamentous networks~\cite{Ronceray:2016}. We thus find
that active stress amplification is rigorously proportional to the
buckling radius $R^*$. We further provide a continuum
counterpart to the ``rope network'' picture previously used to justify
the $r^{1-d}$ decay of stresses in the buckling zone, thus extending
its relevance to non-fibrous materials. Finally, we find that the
external stress $\sigma$ influences stress amplification in an
extremely simple way, as it enters the expressions characterizing the
buckling thresholds, buckling radius and amplification factors only
through the combination $\sigma+\sigma_b$, as illustrated in Fig.~\ref{fig:summary}. More generally, the stress-strain relationship of the medium can be expressed as a relationship between $\sigma+\sigma_b$ and $\gamma+\sigma_b/(d\lambda+2\mu)$ that does not explicitly depend on $\sigma_b$.
In practice, this means that in the fixed-stress ensemble the effect of prestressing the network is identical to that of shifting its buckling threshold $\sigma_b$ by a quantity $\sigma$. In the fixed-strain ensemble, this implies that prestraining the network by $\gamma$ is equivalent to shifting $\sigma_b$ by $\sigma_\text{elastic}=(d\lambda+2\mu)\gamma$.

The results derived in this paper are largely independent on the
detailed characteristics of the elastic material considered, and are
derived without the need of fully specifying a nonlinear stress-strain
relation [see Eq.~\eqref{eq:constitutive}]. Indeed, our only nonlinear
assumption is the plateauing of compressive stresses. Our study however leaves out elastic media with an auxetic linear response,
\emph{i.e.}, exotic materials whose lateral dimension shrinks when
they are compressed vertically. Such materials undergo a
different sequence of buckling transitions, whereby the outside of the
active unit buckles before the inside. The characterization of these
new regimes requires additional assumptions about the material's
nonlinear properties, and generally do not yield closed-form
expressions such as the ones presented here. Finally, our above
discussion focuses on the case where $\sigma_b>0$, \emph{i.e.}, on
materials that dramatically soften under compression. Our results are
nonetheless formally applicable to materials with the opposite
tendency, \emph{e.g.} granular materials that lose all rigidity if
their grains are pulled apart far enough to break the contacts between
them. Indeed, simultaneously reversing the signs of all stresses,
strains and displacements in our study converts the tense radial ropes
underlying the force transmission in the $R_0<r<R^*$ buckling zone of
Fig.~\ref{fig:bucklingstates}(c) into compressed granular columns with
a similar propensity for long-range stress propagation. Whether such a
state can be stable against the lateral buckling of such columns
however remains to be determined.


\begin{acknowledgments}
This work was supported by a PCTS fellowship to PR, the German Excellence Initiative via the program ``NanoSystems Initiative   Munich''   (NIM) and   the Deutsche Forschungsgemeinschaft   (DFG)   via   project   B12   within   the   SFB-1032 to CPB, Marie Curie Integration Grant PCIG12-GA-2012-334053, ``Investissements d'Avenir'' LabEx PALM (ANR-10-LABX-0039-PALM), ANR grant ANR-15-CE13-0004-03 and ERC Starting Grant 677532 to ML. ML's group belongs to the CNRS consortium CellTiss.
\end{acknowledgments}


\begin{thebibliography}{23}
\expandafter\ifx\csname natexlab\endcsname\relax\def\natexlab#1{#1}\fi
\expandafter\ifx\csname bibnamefont\endcsname\relax
  \def\bibnamefont#1{#1}\fi
\expandafter\ifx\csname bibfnamefont\endcsname\relax
  \def\bibfnamefont#1{#1}\fi
\expandafter\ifx\csname citenamefont\endcsname\relax
  \def\citenamefont#1{#1}\fi
\expandafter\ifx\csname url\endcsname\relax
  \def\url#1{\texttt{#1}}\fi
\expandafter\ifx\csname urlprefix\endcsname\relax\def\urlprefix{URL }\fi
\providecommand{\bibinfo}[2]{#2}
\providecommand{\eprint}[2][]{\url{#2}}

\bibitem[{\citenamefont{Jen and Mclntire}(1982)}]{Jen:1982}
\bibinfo{author}{\bibfnamefont{C.~J.} \bibnamefont{Jen}} \bibnamefont{and}
  \bibinfo{author}{\bibfnamefont{L.~V.} \bibnamefont{Mclntire}},
  \bibinfo{journal}{Cell Motil.} \textbf{\bibinfo{volume}{2}},
  \bibinfo{pages}{445} (\bibinfo{year}{1982}).

\bibitem[{\citenamefont{Blanchoin et~al.}(2014)\citenamefont{Blanchoin,
  Boujemaa-Paterski, Sykes, and Plastino}}]{Blanchoin:2014}
\bibinfo{author}{\bibfnamefont{L.}~\bibnamefont{Blanchoin}},
  \bibinfo{author}{\bibfnamefont{R.}~\bibnamefont{Boujemaa-Paterski}},
  \bibinfo{author}{\bibfnamefont{C.}~\bibnamefont{Sykes}}, \bibnamefont{and}
  \bibinfo{author}{\bibfnamefont{J.}~\bibnamefont{Plastino}},
  \bibinfo{journal}{Physiol. Rev.} \textbf{\bibinfo{volume}{94}},
  \bibinfo{pages}{235} (\bibinfo{year}{2014}).

\bibitem[{\citenamefont{Heisenberg and Bella\"iche}(2013)}]{Heisenberg:2013}
\bibinfo{author}{\bibfnamefont{C.-P.} \bibnamefont{Heisenberg}}
  \bibnamefont{and}
  \bibinfo{author}{\bibfnamefont{Y.}~\bibnamefont{Bella\"iche}},
  \bibinfo{journal}{Cell} \textbf{\bibinfo{volume}{153}}, \bibinfo{pages}{948}
  (\bibinfo{year}{2013}).

\bibitem[{\citenamefont{Broedersz and MacKintosh}(2014)}]{Broedersz:2014}
\bibinfo{author}{\bibfnamefont{C.~P.} \bibnamefont{Broedersz}}
  \bibnamefont{and} \bibinfo{author}{\bibfnamefont{F.~C.}
  \bibnamefont{MacKintosh}}, \bibinfo{journal}{Rev. Mod. Phys.}
  \textbf{\bibinfo{volume}{86}}, \bibinfo{pages}{995} (\bibinfo{year}{2014}).

\bibitem[{\citenamefont{Gurtin}(1972)}]{Gurtin:1972}
\bibinfo{author}{\bibfnamefont{M.~E.} \bibnamefont{Gurtin}},
  \emph{\bibinfo{title}{Encyclopedia of Physics}}
  (\bibinfo{publisher}{Springer-Verlag}, \bibinfo{year}{1972}), vol.
  \bibinfo{volume}{VIa/2}, pp. \bibinfo{pages}{1--295}.

\bibitem[{\citenamefont{Ronceray and Lenz}(2015)}]{Ronceray:2015}
\bibinfo{author}{\bibfnamefont{P.}~\bibnamefont{Ronceray}} \bibnamefont{and}
  \bibinfo{author}{\bibfnamefont{M.}~\bibnamefont{Lenz}},
  \bibinfo{journal}{Soft Matter} \textbf{\bibinfo{volume}{11}},
  \bibinfo{pages}{1597} (\bibinfo{year}{2015}).

\bibitem[{\citenamefont{Shokef and Safran}(2012)}]{Shokef:2012}
\bibinfo{author}{\bibfnamefont{Y.}~\bibnamefont{Shokef}} \bibnamefont{and}
  \bibinfo{author}{\bibfnamefont{S.~A.} \bibnamefont{Safran}},
  \bibinfo{journal}{Phys. Rev. Lett.} \textbf{\bibinfo{volume}{108}},
  \bibinfo{pages}{178103} (\bibinfo{year}{2012}).

\bibitem[{\citenamefont{Wang et~al.}(2014)\citenamefont{Wang, Abhilash, Chen,
  Wells, and Shenoy}}]{Wang:2014}
\bibinfo{author}{\bibfnamefont{H.}~\bibnamefont{Wang}},
  \bibinfo{author}{\bibfnamefont{A.~S.} \bibnamefont{Abhilash}},
  \bibinfo{author}{\bibfnamefont{C.~S.} \bibnamefont{Chen}},
  \bibinfo{author}{\bibfnamefont{R.~G.} \bibnamefont{Wells}}, \bibnamefont{and}
  \bibinfo{author}{\bibfnamefont{V.~B.} \bibnamefont{Shenoy}},
  \bibinfo{journal}{Biophys. J.} \textbf{\bibinfo{volume}{107}},
  \bibinfo{pages}{2592} (\bibinfo{year}{2014}).

\bibitem[{\citenamefont{Notbohm et~al.}(2015)\citenamefont{Notbohm, Lesman,
  Rosakis, Tirrell, and Ravichandran}}]{Notbohm:2014}
\bibinfo{author}{\bibfnamefont{J.}~\bibnamefont{Notbohm}},
  \bibinfo{author}{\bibfnamefont{A.}~\bibnamefont{Lesman}},
  \bibinfo{author}{\bibfnamefont{P.}~\bibnamefont{Rosakis}},
  \bibinfo{author}{\bibfnamefont{D.~A.} \bibnamefont{Tirrell}},
  \bibnamefont{and}
  \bibinfo{author}{\bibfnamefont{G.}~\bibnamefont{Ravichandran}},
  \bibinfo{journal}{J. R. Soc. Interface} \textbf{\bibinfo{volume}{12}},
  \bibinfo{pages}{20150320} (\bibinfo{year}{2015}).

\bibitem[{\citenamefont{Rosakis et~al.}(2015)\citenamefont{Rosakis, Notbohm,
  and Ravichandran}}]{Rosakis:2014}
\bibinfo{author}{\bibfnamefont{P.}~\bibnamefont{Rosakis}},
  \bibinfo{author}{\bibfnamefont{J.}~\bibnamefont{Notbohm}}, \bibnamefont{and}
  \bibinfo{author}{\bibfnamefont{G.}~\bibnamefont{Ravichandran}},
  \bibinfo{journal}{J. Mech. Phys. Solids} \textbf{\bibinfo{volume}{85}},
  \bibinfo{pages}{16} (\bibinfo{year}{2015}).

\bibitem[{\citenamefont{Xu and Safran}(2015)}]{Xu:2015}
\bibinfo{author}{\bibfnamefont{X.}~\bibnamefont{Xu}} \bibnamefont{and}
  \bibinfo{author}{\bibfnamefont{S.~A.} \bibnamefont{Safran}},
  \bibinfo{journal}{Phys. Rev. E} \textbf{\bibinfo{volume}{92}},
  \bibinfo{pages}{032728} (\bibinfo{year}{2015}).

\bibitem[{\citenamefont{Ronceray et~al.}(2016)\citenamefont{Ronceray,
  Broedersz, and Lenz}}]{Ronceray:2016}
\bibinfo{author}{\bibfnamefont{P.}~\bibnamefont{Ronceray}},
  \bibinfo{author}{\bibfnamefont{C.}~\bibnamefont{Broedersz}},
  \bibnamefont{and} \bibinfo{author}{\bibfnamefont{M.}~\bibnamefont{Lenz}},
  \bibinfo{journal}{Proc. Natl. Acad. Sci. U.S.A.}
  \textbf{\bibinfo{volume}{113}}, \bibinfo{pages}{2827} (\bibinfo{year}{2016}).

\bibitem[{\citenamefont{Wagner}(1929)}]{Wagner:1929}
\bibinfo{author}{\bibfnamefont{H.}~\bibnamefont{Wagner}}, \bibinfo{journal}{Z.
  Flugtechn. Motorluftschiffahrt} \textbf{\bibinfo{volume}{20}},
  \bibinfo{pages}{8} (\bibinfo{year}{1929}).

\bibitem[{\citenamefont{Davidovitch et~al.}(2011)\citenamefont{Davidovitch,
  Schroll, Vella, Adda-Bedia, and Cerda}}]{Davidovitch:2011}
\bibinfo{author}{\bibfnamefont{B.}~\bibnamefont{Davidovitch}},
  \bibinfo{author}{\bibfnamefont{R.~D.} \bibnamefont{Schroll}},
  \bibinfo{author}{\bibfnamefont{D.}~\bibnamefont{Vella}},
  \bibinfo{author}{\bibfnamefont{M.}~\bibnamefont{Adda-Bedia}},
  \bibnamefont{and} \bibinfo{author}{\bibfnamefont{E.~A.} \bibnamefont{Cerda}},
  \bibinfo{journal}{Proc. Natl. Acad. Sci. U.S.A.}
  \textbf{\bibinfo{volume}{108}}, \bibinfo{pages}{18227}
  (\bibinfo{year}{2011}).

\bibitem[{\citenamefont{Kruse et~al.}(2005)\citenamefont{Kruse, Joanny,
  J{\"u}licher, Prost, and Sekimoto}}]{Kruse:2005aa}
\bibinfo{author}{\bibfnamefont{K.}~\bibnamefont{Kruse}},
  \bibinfo{author}{\bibfnamefont{J.~F.} \bibnamefont{Joanny}},
  \bibinfo{author}{\bibfnamefont{F.}~\bibnamefont{J{\"u}licher}},
  \bibinfo{author}{\bibfnamefont{J.}~\bibnamefont{Prost}}, \bibnamefont{and}
  \bibinfo{author}{\bibfnamefont{K.}~\bibnamefont{Sekimoto}},
  \bibinfo{journal}{Eur. Phys. J. E} \textbf{\bibinfo{volume}{16}},
  \bibinfo{pages}{5} (\bibinfo{year}{2005}).

\bibitem[{\citenamefont{J{\"u}licher et~al.}(2007)\citenamefont{J{\"u}licher,
  Kruse, Prost, and Joanny}}]{Julicher:2007}
\bibinfo{author}{\bibfnamefont{F.}~\bibnamefont{J{\"u}licher}},
  \bibinfo{author}{\bibfnamefont{K.}~\bibnamefont{Kruse}},
  \bibinfo{author}{\bibfnamefont{J.}~\bibnamefont{Prost}}, \bibnamefont{and}
  \bibinfo{author}{\bibfnamefont{J.-F.} \bibnamefont{Joanny}},
  \bibinfo{journal}{Phys. Rep.-Rev. Sec. Phys. Lett.}
  \textbf{\bibinfo{volume}{449}}, \bibinfo{pages}{3} (\bibinfo{year}{2007}).

\bibitem[{\citenamefont{Joanny and Prost}(2009)}]{Joanny:2009}
\bibinfo{author}{\bibfnamefont{J.-F.} \bibnamefont{Joanny}} \bibnamefont{and}
  \bibinfo{author}{\bibfnamefont{J.}~\bibnamefont{Prost}},
  \bibinfo{journal}{HFSP J.} \textbf{\bibinfo{volume}{3}}, \bibinfo{pages}{94}
  (\bibinfo{year}{2009}).

\bibitem[{\citenamefont{Prost et~al.}(2015)\citenamefont{Prost, J\"ulicher, and
  Joanny}}]{Prost:2015}
\bibinfo{author}{\bibfnamefont{J.}~\bibnamefont{Prost}},
  \bibinfo{author}{\bibfnamefont{F.}~\bibnamefont{J\"ulicher}},
  \bibnamefont{and} \bibinfo{author}{\bibfnamefont{J.-F.}
  \bibnamefont{Joanny}}, \bibinfo{journal}{Nat. Phys.}
  \textbf{\bibinfo{volume}{11}}, \bibinfo{pages}{111} (\bibinfo{year}{2015}).

\bibitem[{\citenamefont{Voituriez et~al.}(2005)\citenamefont{Voituriez, Joanny,
  and Prost}}]{Voituriez:2005}
\bibinfo{author}{\bibfnamefont{R.}~\bibnamefont{Voituriez}},
  \bibinfo{author}{\bibfnamefont{J.~F.} \bibnamefont{Joanny}},
  \bibnamefont{and} \bibinfo{author}{\bibfnamefont{J.}~\bibnamefont{Prost}},
  \bibinfo{journal}{Europhys. Lett.} \textbf{\bibinfo{volume}{70}},
  \bibinfo{pages}{404} (\bibinfo{year}{2005}).

\bibitem[{\citenamefont{Bois et~al.}(2011)\citenamefont{Bois, J\"ulicher, and
  Grill}}]{Bois:2011}
\bibinfo{author}{\bibfnamefont{J.~S.} \bibnamefont{Bois}},
  \bibinfo{author}{\bibfnamefont{F.}~\bibnamefont{J\"ulicher}},
  \bibnamefont{and} \bibinfo{author}{\bibfnamefont{S.~W.} \bibnamefont{Grill}},
  \bibinfo{journal}{Phys. Rev. Lett.} \textbf{\bibinfo{volume}{106}},
  \bibinfo{pages}{028103} (\bibinfo{year}{2011}).

\bibitem[{\citenamefont{Kumar et~al.}(2014)\citenamefont{Kumar, Bois,
  J\"ulicher, and Grill}}]{Kumar:2014}
\bibinfo{author}{\bibfnamefont{K.~V.} \bibnamefont{Kumar}},
  \bibinfo{author}{\bibfnamefont{J.~S.} \bibnamefont{Bois}},
  \bibinfo{author}{\bibfnamefont{F.}~\bibnamefont{J\"ulicher}},
  \bibnamefont{and} \bibinfo{author}{\bibfnamefont{S.~W.} \bibnamefont{Grill}},
  \bibinfo{journal}{Phys. Rev. Lett.} \textbf{\bibinfo{volume}{112}},
  \bibinfo{pages}{208101} (\bibinfo{year}{2014}).

\bibitem[{\citenamefont{Chan and Odde}(2008)}]{Chan:2008}
\bibinfo{author}{\bibfnamefont{C.~E.} \bibnamefont{Chan}} \bibnamefont{and}
  \bibinfo{author}{\bibfnamefont{D.~J.} \bibnamefont{Odde}},
  \bibinfo{journal}{Science} \textbf{\bibinfo{volume}{322}},
  \bibinfo{pages}{1687} (\bibinfo{year}{2008}).

\bibitem[{\citenamefont{Lenz}(2014)}]{Lenz:2014}
\bibinfo{author}{\bibfnamefont{M.}~\bibnamefont{Lenz}}, \bibinfo{journal}{Phys.
  Rev. X} \textbf{\bibinfo{volume}{4}}, \bibinfo{pages}{041002}
  (\bibinfo{year}{2014}).

\end{thebibliography}
\end{document}